\documentstyle{mn}

\input{epsf}

\begin{document}

\title[Non-Gaussian bounds in the variance from small
scale CMB observations]{Non-Gaussian bounds in the variance from small
scale CMB observations} 

\author
[E. Gazta\~{n}aga, P. Fosalba, E. Elizalde]
{E. Gazta\~{n}aga, P. Fosalba, E. Elizalde \\
Institut d'Estudis Espacials de Catalunya, Research Unit (CSIC), \\
Edf. Nexus-204 - c/ Gran Capit\'an 2-4, 08034 Barcelona, Spain}

\maketitle

\def\Mpc{{\,h^{-1}\,{\rm Mpc}}}
\def\mpc {h^{-1} {\rm{Mpc}}}
\def\rmd {\rm d}
\def\eg{{e.g., }}
\def\etal{{et~al., }}
\def\spose#1{\hbox to 0pt{#1\hss}}
\def\simlt{\mathrel{\spose{\lower 3pt\hbox{$\mathchar"218$}}
     \raise 2.0pt\hbox{$\mathchar"13C$}}}
\def\simgt{\mathrel{\spose{\lower 3pt\hbox{$\mathchar"218$}}
     \raise 2.0pt\hbox{$\mathchar"13E$}}}
\def\beq{\begin{equation}}
\def\eeq{\end{equation}}
\def\bce{\begin{center}}
\def\ece{\end{center}}
\def\bea{\begin{eqnarray}}
\def\eea{\end{eqnarray}}
\def\ben{\begin{enumerate}}
\def\een{\end{enumerate}}
\def\ul{\underline}
\def\ni{\noindent}
\def\nn{\nonumber}
\def\bs{\bigskip}
\def\ms{\medskip}
\def\wt{\widetilde}
\def\brr{\begin{array}}
\def\err{\end{array}}
\def\dsp{\displaystyle}
\newcommand{\rhobar}{\overline{\rho}}
\newcommand{\xibar}{\overline{\xi}}
\def\Or{{\cal O}}

\font\twelveBF=cmmib10 scaled 1200
\newcommand{\bte}{\hbox{\twelveBF $\theta$}}
\newcommand{\x}{\hbox{\twelveBF x}}
\newcommand{\vv}{\hbox{\twelveBF v}}
\newcommand{\y}{\hbox{\twelveBF y}}
\newcommand{\r}{\hbox{\twelveBF r}}
\newcommand{\k}{\hbox{\twelveBF k}}
\newcommand{\lexp}{\mathop{\bigl\langle}}
\newcommand{\rexp}{\mathop{\bigr\rangle}}
\newcommand{\rexpc}{\mathop{\bigr\rangle_c}{}}
\newcommand{\eq}{{equation~}}

\begin{abstract}

We compare the latest results from CMB experiments
at scales around $l_e \sim 150$ over
different parts of the sky to test the hypothesis that they are
drawn from a Gaussian distribution, as is usually assumed.
Using both the diagonal and the full covariance $\chi^2$ test,
we compare the data with different sets of strategies and
find in all cases incompatibility with the Gaussian hypothesis
above the one-sigma level. We next show how to include 
a generic non-Gaussian signal in the data analysis.
Results from CMB observations can be made compatible with each other
by assuming a non-Gaussian distribution for the signal,
with a kurtosis at a level
$B_4 = \langle{ {\delta_T^4} \rangle}_c /  
 {\langle{ \delta_T^2 \rangle}_c}^2 \simeq 90$. 
A possible  interpretation for this result is that
the initial fluctuations  at the surface of last scattering
are strongly non-Gaussian. 
Another interpretation is that the systematic errors have
been understimated in all observations by a factor of two. 
Other explanations include foreground contamination, non-linear 
effects or a combination of them.

\end{abstract}


\section{Introduction}

A basic ingredient to understand the formation of 
large scale structures in our Universe is the distribution of
initial conditions. Have fluctuations been generated 
in the standard  inflationary epoch or do they
require topological defects or more exotic assumptions
for the initial conditions?
While the former assumption
typically produces a Gaussian distribution
(Bardeen, Steinhardt, Turner 1983) the latter  
involves strong non-Gaussianities (\eg Vilenkin 1985,
Turok \& Spergel 1991).
This issue can be addressed both in the
present day Universe fluctuations, as traced by the galaxy
distribution (\eg Silk \& Juszkiewicz 1991, 
Gazta\~naga \& M\"ah\"onen 1996), 
or in the the anisotropies of the
cosmic microwave background (CMB) (\eg Coulson \etal 1994, 
Smoot \etal 1994).  Here we will address the latter possibility 
in a somewhat indirect way.
One important contribution to the uncertainties in the 
measurements of the amplitude of the CMB  comes from the sample  
variance. That is, the uncertainty due to the finite size of
the observational sample.
In order to estimate these sampling errors it 
is common practice to assume that the underlying 
signal is Gaussian (\eg Bond \etal 1994).
 These errors are added to other sources of 
error to test models of structure formation or to compare between
experiments. A non-Gaussian signal can produce different 
sampling errors, and this possibility has already been proposed as 
a way to reconcile the discrepancies between different
experiments (Coulson \etal 1994, Luo 1995)

Here we propose to go a step further and use the estimated
discrepancies or variance between different experiments to 
place bounds on the degree of non-Gaussianity. In order to do
this we will assume that the quoted systematic errors in each
experiment are accurate, at least on average.
We will focus on results which are either from different parts 
of the sky or, when over the same area, from multipoles with 
windows that are well separated
appart. Our strategy is not to average results from a given experiment, 
but to find as many  independent results as possible in order to
have a large sampling over the underlying distribution.

\section{Sample Variance}
\label{sample}

We want to study the sample variance of CMB experiments over independent 
sky regions or subsamples. We will denote the ensemble average
by $\langle \cdots \rangle$. In each subsample we have measurements on
several resolution cells or
patches, whose averages (within the subsample) we
denote by bars. As usual, we assume the {\em fair sample hypothesis} and in
particular that the
ensemble averages can be identified with spatial averages
(\S30 Peebles 1980).  
In order to derive the sample variance of the temperature 
fluctuations in the sky for a generic non-Gaussian field, 
we define our radiation field as 
$\Delta_m = T_m - \overline{T}$,
with $T_m$ the temperature field at a point within certain
patch $m$ over which we calculate the subsample average, 
$\overline{T}$. 
Notice that the normalized field is given by
$(\delta_T)_m = {{T_m - \overline{T}}\over{\overline{T}}}$.
According to this notation,
all magnitudes derived from the field $\Delta_m$ may have 
dimensions. 
It follows from its definition that the subsample average  
$\overline{\Delta}=0$, so that its
variance is:
\beq
\overline{\Delta^2} = {1 \over N} \sum_m^N {\Delta_m^2} 
\eeq
In the literature this  quantity is denoted by $\delta T^2$, which 
should not be confused with our notation for 
the dimensionaless local fluctuation $\delta_T$.
The sample variance of $\overline{\Delta^2}$
is therefore the variance of the variance of the temperature field 
\bea
 Var \left( \overline{\Delta^2} \right) &=&
\langle \left( \overline{\Delta^2} 
- {\langle  \overline{\Delta^2} \rangle} \right)^2  \rangle 
= {1 \over N^2}\, \left\{ 
\sum_m { \langle \left( \Delta_m^2 
- \langle \Delta_m^2 \rangle \right)^2  \rangle} \, \right. \nonumber \\
 &+& \left. 
\, \sum_{m \neq n} { \langle \left(  {\Delta_m^2} 
-  \langle \Delta_m^2 \rangle \right)
\left(  {\Delta_n^2} 
-  \langle \Delta_n^2  \rangle  \right) \rangle }  \right\}, \nonumber
\eea
where we have made use
of the fact that the sum commutes with the ensemble average,
$\langle \cdots \rangle$. We 
furthermore assume that the subsamples are large enough to
neglect the average cross correlations between patches as compared to the
mean square contributions (see Fig.1 in  Scott et al 1994).
We can then drop the last term and rewrite the first one
by commuting back the sum and the averages:
\bea
\label{ngsv}
Var \left( \overline{\Delta^2} \right)
&=& {1 \over N^2}\, \langle
\sum_m  \left( \Delta_m^4 
- 2 \Delta_m^2 \, \langle \Delta^2 \rangle + \langle \Delta^2 \rangle^2
\right)  \rangle \, \nonumber \\
&=& {1 \over N}\, \left\{ {\langle {\Delta^4} \rangle}_c + 
2\,{\langle {\Delta^2} \rangle}_c^2 \right\} ,
\label{sv}
\eea
where we have used that for any $X$: 
 $\langle \bar{X}  \rangle =\langle X \rangle =\langle X_m \rangle $,
and we have applied the standard definition for $\langle...\rangle_c$,
connected moments or cumulants (\eg Kendall, Stuart \& Ort 1987).


Throughout the analysis
we shall consider a general family of 
non-Gaussian signals with {\it dimensional} scaling,
which is choosen
because it enters at the same level than the Gaussian contribution 
in the sample variance (see Discussion). 
For {\it dimensional} scaling, we have that the 4th order cumulant
scales with the square of the 2nd order cumulant, so that:
\beq
B_4 \equiv {{\langle {\delta_T}^4 \rangle}_c \over{
 {{\langle {\delta_T}^2 \rangle}_c}^2 }}
= {{\langle {\Delta }^4 \rangle}_c \over{
 {{\langle {\Delta }^2 \rangle}_c}^2 }},
\eeq 
is a constant
(\eg independent of  ${\langle {\Delta }^2 \rangle}_c$).
In terms of $B_4$, expression (\ref{sv}) then reads
\beq
Var(\delta T^2) \equiv  Var \left( \overline{\Delta^2} \right) =
{1\over N}\, \left(2\,+\,B_4 \right)\, 
{\langle \Delta^2 \rangle}_c^2.
\label{vardelta}
\eeq
The Gaussian sample variance corresponds to the particular case $B_4 =0$.
It is important to stress the general applicability of (\ref{sv})
for non-Gaussian processes. 

\section{Small-scale CMB Data compilation}

\begin{table*}

\caption[junk]{Available small-scale experimental data 
within the range $90 \leq l_e \leq 200$.
The superscript $a$ denotes 
the MAX experiments (see Tanaka et al. 1995, and references therein). 
They are labeled according to the sky patch and flight, 
MAX-{\em Sky patch}\, ({\em flight}); $b$ denotes the two MSAM1-94
(single difference) experiments reported in Cheng et al. 1996,
referring to independent sky regions in RA; $c$ denotes the
$Saskatoon '95$ experiments, where
SK95C$n$ 
correspond to the $95$ 
CAP region, SK94K$n$ and
SK94Q$n$ 
to the K and Q band experiments in the $ '94$ flight, with
the $n$ point chopping
strategy in each case (see Netterfield et al 1997);
$d$, 
(see de Bernardis et al. 1994); 
$e$ denotes PYTHON-III$_L$ and PYTHON-III$_S$ for the subtractive large 
and small chop-window measurements, respectively (see, Platt et al. 
1996).}

\begin{center}

\begin{tabular}{|r|l||c|c|c|c|c|c|c||}
\hline
used in Test & CMB experiments & ${\delta T_l} \ (\mu K)$  & $ \sigma_{\delta T}^G \ (\mu K)$
& $\sigma_{sv}^G \ 
(\mu K)$ &
$l_e$ & $+ \Delta l_e$ & $- \Delta l_e$ & $ N_{points}$   \\
\hline 
 A1,A2,B & $^a$MAX-GUM (3) & 78 & 18 & 9 & 145 & 100 & 60 & 39 \\ \cline{3-9}
 A1,A2,B&$^a$MAX-MUP (3)     & 26 & 10 & 3 & 145 & 100 & 60 & 39 \\ \cline{3-9}
 A1,A2,B&$^a$MAX-ID (4)      & 46 & 18 & 7 & 145 & 100 & 60 & 21 \\ \cline{3-9}
 A1,A2,B&$^a$MAX-SH (4)      & 49 & 19 & 8 & 145 & 100 & 60 & 21 \\ \cline{3-9}
 A1,A2,B&$^a$MAX-HR5127 (5)  & 33 & 15 & 4 & 145 & 100 & 60 & 29 \\ \cline{3-9}
 A1,A2,B&$^a$MAX-PH (5)      & 52 & 15 & 7 & 145 & 100 & 60 & 29 \\ \cline{3-9}
A1,A2,B&$^b$MSAM-2 beam(1) & 61 & 37 & 7 & 159 & 75 & 76 & 34 \\ \cline{3-9}
A1,A2,B&$^b$MSAM-2 beam(2)  & 28 & 18 & 3 & 159 & 112 & 82 & 34 \\ \cline{3-9}
A2,B&$^c$SK95C6      & 64 & 17 & 7 & 135 & 6 & 37 & 48 \\ \cline{3-9}
A2,B&$^c$SK95C7	& 72 & 19 & 7 & 158 & 7 & 38 & 48 \\  \hline
B & $^c$SK95C5		 & 54 & 17 & 8 & 108 & 8 & 32 & 24 \\ \cline{3-9}
B&$^c$SK95C8 		& 81 & 20 & 8 & 178 & 6 & 38 & 48 \\ \cline{3-9}
B&$^c$SK95C9 		& 76 & 20 & 8 & 197 & 8 & 37 & 48 \\ \cline{3-9}
B&$^c$SK94K5 		& 44 & 14 & 6 & 96 & 21 & 19 & 24 \\ \cline{3-9}
B&$^c$SK94K6 		& 33 & 15 & 3 & 115 & 21 & 19 & 48 \\ \cline{3-9}
B&$^c$SK94Q9  	& 138 & 55 & 14 & 176 & 20 & 23 & 48 \\ \cline{3-9}
B&$^d$ARGO	& 39 & 6 & 3 & 98 & 70 & 38 & 63 \\ \cline{3-9}
B&$^e$PYTHON-III$_L$ & 59 & 14 & 3 & 178 & 61 & 45 & 158 \\ \cline{3-9}
B&$^e$PYTHON-III$_S$ & 54 & 14 & 3& 92 & 7 & 39 & 127 \\ 
\hline 
\end{tabular}

\label{cmbexp}
\end{center}

\end{table*}

For each CMB experiment over a given subsample,
labeled $i$, we denote as usual
$\delta T(i) \equiv \sqrt{ \left(\overline{\Delta^2} \right)_i}$  
ie, $\delta T(i)$ is the {\em rms} temperature anisotropy,
from which one estimates the band power $\delta T_l (i) $ for every $l$ 
multipole
component of the power spectrum.
Table \ref{cmbexp} shows a compilation of available 
 data from small-scale experiments
for scales within the range $90 \leq l_e \leq 200$. This interval
is specially suitable for a $\chi^2$ analysis since
it is the most densely sampled, according to observational reports. The
scale and size of each window peaks at multipole
number $l_e$  and  has a width given by the  $\pm \Delta l_e$
interval (computed as the scales at which the window falls
to a factor of $e^{-0.5}$ of the peak value). 
Each input in this table 
corresponds to independent sky patches or well separated windows. 
The total quoted error, $
\sigma_{\delta T}^G$, includes the calibration uncertainty,
the sampling and the instrumental errors.
The number of independent points
for the statistical analysis is given by the independent bins in RA 
in each observation. The data esentially follows Ratra \etal 1997,  
but several cases are taken from the original 
observational reports.
Notice that performing the correct window weighting of
the CMB models (\eg Ratra et al. 1997) hardly changes the final
results within the errors, and therefore the   
flat band hypothesis that we are using for comparing 
individual experiments should be equally accurate.
 The data points with their
errors (horizontal ones corresponding to the window width) are displayed
in Figure 1.

\begin{figure} 
\centering
\centerline{\epsfysize=8.truecm 
\epsfbox{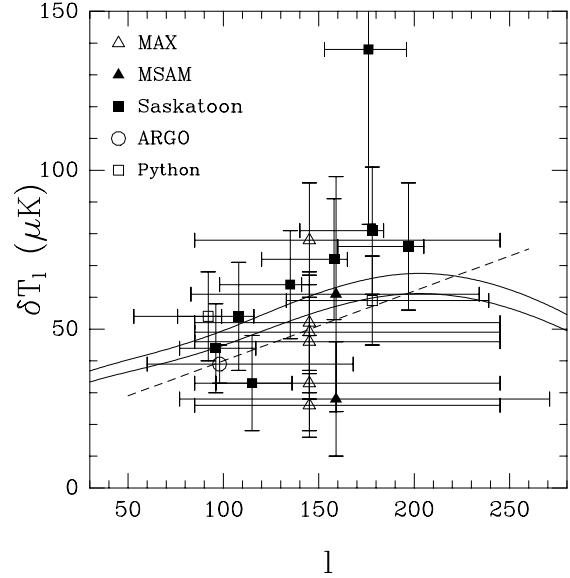}}
\caption[junk]{Band power estimates of
the {\em rms} temperature anisotropy $\delta T_l$
for observations given in Table 1.
The vertical error bars show the (symmetrized) total errors in $\delta T_l$
 while the horizontal ones stand for the width
of the windows. The dashed line
is the best fit slope to the data, $\delta T_l = (11/50) l_e + 18$.
Continuous lines show the standard CDM model for two normalizations: 
$Q_{rms} = 20 \mu$ K (top), and $18 \mu$ K (bottom).}
\label{cmbchi}\end{figure}

In the above results,  
Gaussian (G) statistics have been assumed to calculate the sampling variance 
contribution to the error
and this is always included in the quoted errors.
For a general non-Gaussian case
 we would like to replace this contribution
with the more general expression given above.
The sampling variance estimation in the experiments
 is usually done with Gaussian Monte-Carlo simulations.
Here, to estimate
this contribution we will use its theoretical
expectation. We first write the rms 
error $\sigma_{sv}$ from the
sampling variance as: 
$\sigma_{sv}[\delta T^2]= 2 ~\delta T~\sigma_{sv}[\delta T]$,
so that we find from
equation (\ref{vardelta}) that:
\beq
\sigma_{sv}[\delta T] = \sqrt{ {1\over 2 N}\,\left(1+{B_4\over{2}}\right)}
 ~\delta T ,
\label{sigmasv}
\eeq
were $N$ is the number of independent observations. 
In this equation, we have 
used the individual experiment (or subsample) averages 
$\delta T$
instead of the {\em ensemble} averages: ${\langle \Delta^2 \rangle}^{1/2}_c$. 
This is not exact, but reproduces
better what is done in each experiment to estimate the Gaussian
sampling error,  which we denote $\sigma^G_{sv}$ (i.e., for $B_4=0$ above).
We have checked that the results shown below do not
significantly depend on such approximation. 
We then assume that the {\it total} error for a Gaussian signal $\sigma^G_{\delta T}$,
given in the observational reports,
can be obtained by adding in quadrature $\sigma^G_{sv}$
 to the other errors 
(\eg the instrumental and calibration errors).
The values of $\sigma^G_{\delta T}$ and  $\sigma^G_{sv}$ 
for each experiment are 
shown in Table \ref{cmbexp}.

We next carry out a Chi-square analysis taking different number of points
according to the following:
\begin{itemize}
\item{{\bf Test A:} Taking a band as narrow 
and densely sampled as possible, so
that we can neglect any dependence of the signal with the scale, $l$.
We consider two cases:
{\bf A1, A2}: which correspond to the first $8 \& 10$ 
points of Table \ref{cmbexp}, respectively.}

\item{{\bf Test B:} Taking a wider band, as densely sampled as possible, 
and computing the
$\chi^2$ value with ({\bf B1}) and without ({\bf B2})
a linear fit to the signal with $l_e$, i.e.
removing a possible scale
dependence of the power spectrum in the analysis.}
\end{itemize}

The $\chi^2$ values are to be obtained from the 
full covariance analysis:

\beq
\chi^2(c_{ij}) = \sum_{ij} \, D_i \, C_{ij}^{-1} D_j,
\label{covar}
\eeq
were $D_i\equiv \, \langle \delta T_l \rangle -\delta T_l (i)$
are differences between individual observations (in Table \ref{cmbexp})
and a theoretical mean value, $ \langle \delta T_l \rangle$,
which in general  varies with $l$, and $C_{ij} = <D_i D_j>$ is the corresponding
covariance matrix. The diagonal terms 
 $C_{ii}= \sigma_{\delta T}^2(i)$,
 are the individual errors in Table \ref{cmbexp}. 
When the $D_i$'s  are independent, the covariance matrix
becomes diagonal:
\beq
\chi^2(c_{ii}) = \sum_i D_i^2 \, C_{ii}^{-1} =
\sum_i {\left( \langle \delta T_l \rangle -\delta T_l (i) 
\over \sigma_{\delta T} (i) \right)^2} ,
\label{diag}
\eeq
The mean
$\langle \delta T_l \rangle$ in each test is estimated from the individual
values $\delta T_l (i)$
weighted by the inverse of the
variance $\sigma_{\delta T}^2(i)$, which produces a  minimum
$\chi^2$. 
We have done both a diagonal (\ref{diag}) and a
full covariance
analysis (\ref{covar}), taking into account 
the correlations due to 
calibration uncertainties and the overlap of the window functions.
For the off-diagonal terms we use the following:
\beq
C_{ij} =  \kappa_{ij}^2  \, \langle \delta T_l \rangle \,
 \langle \delta T_{l'} \rangle
 + w_{ij} \, \sigma_i^{sv}\sigma_j^{sv} ~~~~ i \neq j
\label{cij}
\eeq
The first term correspond to the calibration error, 
where $\kappa_{ij}$ is zero for observations
$i$, $j$ in different experiments, and otherwise  
is the $\%$ rms 
calibration error of the corresponding
experiment ($\kappa_{ij}=0.14,0.10,.20,0.10,0.05$ 
for Saskatoon, MAX, Phyton, MSAM
and ARGO respectively). 
Note that we need different subscript, $l$ and $l'$, in (\ref{cij})
because the mean can change with $l$.
The second term corresponds to the overlap of the window functions,
were $\sigma_i^{sv}$ are the sampling variance
errors (third column of Table 1) and $w_{ij}$ is 
 the normalized overlap
between the windows (estimated using $l_e \pm \Delta l_e$) 
but only when $i$ and $j$
are sampling the same patch of the sky, otherwise is zero. 
%
 When the mean is definned with the
same data, rather than with a theory, the off diagonal terms tend
to cancel out, as $D_i$ fluctuates around the mean, but in general
the cross-correlations could either increase or decrease
the final $\chi^2$.

\begin{table*}

\caption{$\chi^2$ analysis of combined experiments.}

\begin{center}

\begin{tabular}{|c||c|c|c|c|c|c|c|c|c|}
Test & \multicolumn{4}{c|}{Experiments} & $\chi^2(c_{ii})$/DOF  & $P(\chi^2)$ & 
$\langle \delta T_l \rangle (\mu K)$ &  $B_4$  \\ \hline
\hline 
{\bf A1} &\multicolumn{4}{c|}{MAX+MSAM} & 8.4/5 & 0.14 & 40.7 & $16-238$
\\ \hline
{\bf A2}&  \multicolumn{4}{c|}{
MAX+MSAM+SK95C6 \& 7} & 12.2/7 & 0.09  & 
45.1 & $34-276$
\\ \hline 
& Experiments & $\chi^2(w_{ij})$ & $\chi^2(\kappa_{ij})$ &  $\chi^2(c_{ij})$ & & & & 
\\ \hline
{\bf B1} &  ALL & 22.5 & 24.2 & 23.2 & 23.9/16 & 0.09  & 46.3 & $22-140$
\\ \hline
{\bf B2} & ALL+slope & 19.2 & 19.9 & 19.7 & 20.0/15 & 0.17 & $0.22 l_e +18$ 
& $31-171$ \\ \hline 
{\bf T1} & ALL+CDM1 & 21.9 & 23.0 & 22.8 & 22.4/16 & 0.12 & $C_2=18 \mu K$
& \\ 
{\bf T2} & ALL+CDM2 & 29.0 & 29.7 & 29.6 & 29.2/16 & 0.02 & $C_2=20 \mu K$
& \\ 
\hline 
\end{tabular}

\label{chi}
\end{center}

\end{table*}

 Table \ref{chi} displays the 
$\chi^2$ 
values for all the cases involved in Test {\bf A} and {\bf B}. 
DOF denotes the number of degrees of freedom: $N-3$ (two parameters
are correlated to the data: the mean and $B_4$)
for all cases, except for the last case where DOF$=N-4$ ---since the
slope incorporates one extra parameter to the computation. 
The values of the $\chi^2$, its probability $P(\chi^2)$ and
 $\langle \delta T_l \rangle$ shown
in Table \ref{chi} correspond to the Gaussian case, $B_4=0$,
and $\sigma_{\delta T}=\sigma_{\delta T}^{G}$.
In the last test we find that the 
linear relation that minimizes the $\chi^2$ is:
$\langle \delta T_l \rangle= (11/50)~ l_e ~+ 18$.
All cases considered show a disagreement with 
the Gaussian hypothesis 
above the $1 \sigma$ level, \eg $P(\chi^2)<0.33$, and close
to the  $2 \sigma$ level of significance,  \eg $P(\chi^2) \simeq 0.05$.

The values of $\chi^2(c_{ii})$ in Table \ref{chi} correspond to the
diagonal analysis (\ref{diag}).
The full covariance analysis including both terms
in (\ref{cij}) is given by $\chi^2(c_{ij})$,
while $\chi^2(k_{ij})$ 
and $\chi^2(w_{ij})$ only takes into account the
first or second terms in (\ref{cij}), respectively.
The window overlap, although significant in some of the Saskatoon
values, hardly makes any difference overall.
Thus the full  covariance
analysis, $\chi_2(c_{ij})$,  differs from the
diagonal one, $\chi_2(c_{ii})$,
by less that $3\%$, which hardly changes the significance of
the analysis below. Therefore, we shall concentrate on the
diagonal analysis (\ref{diag})
 to take advantage of is simplicity.

For a non-Gaussian signal, 
the total error above, $\sigma_{\delta T}$, should 
include the sampling variance for the corresponding
non-Gaussian distribution, in equation (\ref{sigmasv}),
as well as the instrumental
and calibration errors. This can be simply related to the {\it total}
(Gaussian) error $\sigma_{\delta T}^{G}$, quoted in Table \ref{cmbexp},
by :
\beq
\sigma_{\delta T}(i)  = \sigma_{\delta T}^{G}(i)  \, \left\{ 1+
{{B_4}\over{8 N_i^2 \left[\sigma_{\delta T}^{G} (i) /\delta T_l (i) 
\right]^4}} \right\}^{1/2},
\eeq
which reduces to (\ref{sigmasv}) when there are no systematic errors:
$\sigma_{\delta T}^{G}=\sigma_{sv}^{G}$.
The range for the
non-Gaussian  parameters $B_4$ shown in  Table \ref{chi}
are the values needed to produce a ${\chi^2}$ value
corresponding to an interval
of confidence between $25\%$-$75\%$.
This range narrows as the number of
data points increases, but the mean values are always away from zero.

Note that our approach is not totally consistent.
We are assuming a non-Gaussian distribution for the signal but
we determine the confidence intervals using the $\chi^2$ 
distribution, which assumes a Gaussian likelihood.
The whole analysis improves substantially by repeating it in terms of a 
non-Gaussian likelihood function. In the limit $\chi^2 \gg N$
and small variance, it is possible to relate the Gaussian 
confidence intervals with
the corresponding non-Gaussian ones in terms of 
$B_4$ and $B_3^2$, where
$B_3 = \langle{ {\delta_T^3} \rangle}_c /  
 {\langle{ \delta_T^2 \rangle}_c}^{3/2}$ (see Amendola 1996).
Within the limitation that $\chi^2 \gg N \gg 1$ (which
restricts applicability to Test {\bf B} only), it turns out 
that the confidence intervals obtained above are widened ---when the 
non-Gaussian corrections are taken into account--- by a factor 
between 1.2 and 2.
To do this estimation one has to assume something
for the value of $B_3^2$.
The later factors 
correspond to $B_3=0$ and $B_4$ in Table \ref{chi},
 which is the most conservative case.
This increases the significance to 
well above $2\sigma$ for both of the Test {\bf B}  cases in Table \ref{chi}.
Fosalba \etal (1997) have found that typical values of $B_3$ in 
several non-Gaussian distributions with {\it dimensional} scaling
lie just below $B_3^2 = B_4$. 
For each assumed value of $B_3$ we can now find in a consistent
way the values of $B_4$ needed in the $\chi^2$ to get
 an interval
of confidence between $25\%$-$75\%$ in the non-Gaussian likelihood.
 For  $|B_3| =8 \simlt B_4^{1/2}$, the allowed ranges for $B_4$ 
corresponding to Test {\bf B1} and {\bf B2} on Table \ref{chi}
shrink to $B_4=70-90$ and $B_4=90-130$.
The improvement in this case is quite 
remarkable, but the range of allowed 
values of $B_4$ increases as $B_3$ approaches zero.

\section{Discussion}
\label{discuss}

Our analysis shows that using the quoted error bars
in different CBM experiments, for scales within the range
$90 \leq l_e \leq 200$, the Gaussian hypothesis can be
rejected at a 91\% or 83\% level, depending on the test
(see Table 2). This is not a very significant level,
and may also be regarded as 9\% or 17\% agreement.
Our main point is  to take these results at face value 
and see what can be said about 
non-gaussianities if we seek for a better agreement.
We have shown how this can be done for a general class of non-Gaussian
distributions by predicting the sampling error. Observations can be made 
compatible with each other at a 1-sigma level, if the
distribution for the signal has a kurtosis at a level of
$B_4 = \langle{ {\delta_T^4} \rangle}_c /  
 {\langle{ \delta_T^2 \rangle}_c}^2 \simeq 90$. 
We have repeated the whole analysis taking out each experiment one by one,
showing that this result is not dominated by a  single measurement.
We have also considered subsets of separated experiments (\eg Test {\bf A1, A2})
and find that this conclusion is robust.
As the mean signal is defined from the data
to obtain the mimimum $\chi^2$, 
any comparison with models
could only lead to a more significant disagreement.
We have also used the code of Seljak \& Zaldarriaga (1996) 
for different CDM universes as the input shape for the mean signal 
in the $\chi^2$. We normalized  the amplitudes according to a
quadrupole  $C_2=18 \mu K$, Test {\bf T1}, or
$C_2=20 \mu K$, Test {\bf T2}, as suggested by COBE observations
(\eg Bennet, \etal 1996). These two normalizations of the
 standard CDM models are shown as the upper 
and lower continuous lines in Figure 1.
The $\chi^2$ and $P(\chi^2)$ values are shown in Table 2.
This illustrates that adding curvature to the input model
does not reduce the $\chi^2$ values; the linear model
is a good approximation to the CDM models 
for this narrow range of $l$.


A possible  interpretation for this result is that
the initial fluctuations  at the surface of last scattering
are strongly non-Gaussian. Even if this initial distribution were
purely Gaussian, it is not clear yet how
non-linear effects in the CMB fluctuations or reionization
(\eg Dodelson \& Jubas 1995) would change the final observed
distributions, although the calculations for some of 
the relevant effects indicate only mild deviations with 
cumulants of {\it hierarchical} type
 (see \eg Mollerach \etal 1995, Munshi \etal 1995).
Another interpretation is that the systematic errors have
been understimated. If we artificially double the systematic errors
in {\it all} experiments at once, we find $\chi^2= 13$ for
Test {\bf B1}, which indicates an agreement at the 67\%
confidence level. Other possibilities include
foreground contamination, which could be in the form of large spots
that should typically induce non-Gaussian fluctuations (although
de Oliveira-Costa \etal 1997, found this  contamination to reduce
the Saskatoon normalization by only 2\%).

Besides the {\it dimensional} scaling  ${{\langle{ {\delta_T}^4} \rangle}_c 
= B_4 {{\langle{ {\delta_T}^2} \rangle}_c^2}}$,
we have also considered another family of non-Gaussian models: 
the case of the {\it hierarchical} scaling mention above,
where  ${{\langle{ {\delta_T}^4} \rangle}_c 
= S_4 {{\langle{ {\delta_T}^2} \rangle}_c^3}}$.
A similar analysis for the {\it hierarchical} scaling, yields:
$S_4 \simeq 10^{12}$.
As the variance $\delta T^2$ is of the same order in all data,
$B_4$ just parametrizes $\langle{ {\delta_T^4} \rangle}_c$
 for any non-Gaussian distribution
and the above value agrees well with 
the na\"{\i}ve expectation: 
$S_4 \simeq B_4/\delta T^2 \simeq B_4 \times 10^{10}$.
Within the large parameter space for
non-Gaussian distributions, the values we find
for $B_4$ and $S_4$ lie in the strongly non-Gaussian cases.
In typical non-Gaussian
models with {\it hierarchical} scaling one has
$S_4 \simlt 10^2$ (\eg Fosalba \etal 1997), 
much smaller than our result $S_4 \simeq 10^{12}$.
For matter fluctuations, $\delta_m$,
 gravitational growth from Gaussian initial 
conditions also gives
$S_4$ of order $10-100$ (\eg Bernardeau 1994).
For non-Gaussian initial conditions, the topological defects 
from phase transitions, like textures (\eg Turok \& Spergel 1991),
predict $B_4 \simeq 1$, for $\delta_m$, and have been 
measured around this value 
in N-body simulations (Gazta\~naga \& M\"ah\"onen 1996), while in the
 present study we get values around $B_4 \simeq 90$.
Thus, the estimated amplitudes for $B_4$ and $S_4$ seem to indicate
high levels of non-Gaussianity, at least according to $\delta_m$
standards. Note that, in principle, one would expect lower levels
of non-Gaussianity in $\delta_T$ than in $\delta_m$, as
the former comes as an integrated effect, at
least on  large scales (see Scherrer $\&$ Schaefer 1995).
The large difference in the order of magnitudes between  $S_4\simeq 10^{12}$
and $B_4\simeq 10^{2}$ indicates that 
{\it dimensional} scaling is a more adequate representation for our
findings  than the {\it hierarchical} scaling. 
If this is the case one would also
typically expect a non-vanishing value for $B_3$ of order $B_3^2 \simlt B_4$ 
Thus,  we would have $|B_3| \simlt 10$ or $|S_3| \simlt 10^6$, 
much larger than the sampling variance expected in Gaussian models,
$\Delta |B_3| \simeq 1$ (\eg Srednicki 1993). For $B_3^2 \simeq B_4$, 
we can put a very tight constraint on $B_4$
to lie between  $B_4=70-90$, for a flat power spectrum, or
$B_4=90-130$ for the best fitted slope in Table \ref{chi}.

\section*{Acknowledgements}

E.G. and P.F. wish to thank Scott Dodelson for encouraging discussions
and the Fermilab group for their kind hospitality during our visit in 
the fall 1996,  were this project
started. E.G. acknowledges support from CIRIT (Generalitat de
Catalunya) grant 1996BEAI300192. 
This work has been
supported by CSIC, DGICYT (Spain), project
PB93-0035, and CIRIT, grant GR94-8001.

\section{References}

\def\refe {\par \hangindent=.7cm \hangafter=1 \noindent}
\def\aj {ApJ,}
\def\na {Nature,}
\def\aa {A\&A,}
\def\prl {Phys.Rev.Lett.,}
\def\prd {Phys.Rev. D,}
\def\phr {Phys.Rep.}
\def\ajs{ApJS,}
\def\mn {MNRAS,}
\def\apl {Ap.J.Lett.,}

\refe Amendola, L., 1996, Astro. Lett. and Comm. 33, 63
\refe Bardeen, J.M., Steinhardt, P.J., Turner, M.S., 1983 \prd 28 679
\refe Bennet, C.L. \etal, 1996, \apl 464, 1 
\refe Bernardeau, F.,  1994 \aa 291, 697
\refe Bond, J.R., Crittenden, R., Davis, R.L., Efstathiou, G., 
Steinhardt, P.J., 1994, \prl 72, 13
\refe Cheng, E.S., \etal 1996, \apl 456, 71
\refe Coulson, D., Ferreira, P., Graham P., Turok, N., 1994, \na 368, 27
\refe de Bernardis P., \etal 1994, \apl 422, 33
\refe de Oliveira-Costa, A., \etal 1997, astro-ph/9705090 
\refe Dodelson, S., Jubas, J.M., 1995 \aj 439, 503
\refe Fosalba, P., Gazta\~{n}aga, E., Elizalde, E., 1997, in preparation
\refe Gazta\~naga, E., M\"ah\"onen, P., 1996, \apl 462, 1
\refe Kendall, M., Stuart, A. Ort, J.K., 1987, 
Kendall's Advanced Theory of Statistics,
Oxford University Press 
\refe Luo, X., 1995, \aj 439, 517
\refe Netterfield, C.B., Devlin, M.J., Jarosik, N., Page, L., Wollack, E.J.,
1997, \aj 474, 47
\refe Mollerach, S., Gangui, A., Lucchin, F., Matarrese, S., 1995, \aj 453, 1 
\refe Munshi, D., Souradeep, T., Starobinsky, A., 1995, \aj 454, 556
\refe Platt, S.R., Kovac, J., Dragovan, M., Peterson, J.B., Ruhl, J.E., 
1996, astro-ph/9606175
\refe Ratra, B., Sugiyama, N., Banday, A.J., G\'{o}rski, K.M., 
1997, \aj 481, 22 
\refe Scherrer, R.J., Schaefer R.K., 1995, \aj 446, 44
\refe Scott, D., Srednicki, M., White, M., 1994 \apl 421, 5
\refe Seljak, U., Zaldarriaga, M., 1996, \aj, 469, 437
\refe Silk J., Juszkiewicz, R. 1991, \na 353, 386 
\refe Smoot, G.F., Banday, A.J., Kogut, A., Wright, E.L., Hinshaw, G.,
Bennett, C.L. 1994, \aj 437, 1
\refe Srednicki, M., 1993, \apl 416, 1
\refe Tanaka, S.T. \etal 1995, astro-ph/9512067
\refe Turok N., Spergel, D.N., 1991 \prl 66, 3093 
\refe Peebles, P.J.E., 1980, {\it The Large Scale Structure of the 
Universe:} Princeton University Press
\refe Vilenkin, A., \phr 121, 263

\end{document}